\documentclass[letterpaper]{article} 

\usepackage{aaai24}  
\usepackage{times}  
\usepackage{helvet}  
\usepackage{courier}  
\usepackage[hyphens]{url}  
\usepackage{graphicx} 
\usepackage{natbib}  
\usepackage{caption} 
\usepackage{algorithm}
\usepackage{algorithmic}

\usepackage{booktabs}
\usepackage{multirow}
\usepackage{color}
\usepackage{comment}

\usepackage{newfloat}
\usepackage{listings}
\DeclareCaptionStyle{ruled}{labelfont=normalfont,labelsep=colon,strut=off} 
\floatstyle{ruled}
\newfloat{listing}{tb}{lst}{}
\floatname{listing}{Listing}
\title{AAAI Press Formatting Instructions \\for Authors Using \LaTeX{} --- A Guide}

\title{LLM-Powered Code Vulnerability Repair \\with Reinforcement Learning and Semantic Reward}
\author {
    Nafis Tanveer Islam\textsuperscript{\rm 1},
    Joseph Khoury\textsuperscript{\rm 2},
    Andrew Seong\textsuperscript{\rm 1},
    Mohammad Bahrami Karkevandi\textsuperscript{\rm 1},
    Gonzalo De La Torre Parra\textsuperscript{\rm 3}, 
    Elias Bou-Harb\textsuperscript{\rm 2},
    Peyman Najafirad\textsuperscript{\rm 1}
}
\affiliations {
    \textsuperscript{\rm 1}University of Texas at San Antonio\\
    \textsuperscript{\rm 2}Louisiana State University\\
    \textsuperscript{\rm 3}University of the Incarnate Word
    
}

\usepackage{bibentry}

\begin{document}

\maketitle

\begin{abstract}
In software development, the predominant emphasis on functionality often supersedes security concerns, a trend gaining momentum with AI-driven automation tools like GitHub Copilot. These tools significantly improve developers' efficiency in functional code development. Nevertheless, it remains a notable concern that such tools are also responsible for creating insecure code, predominantly because of pre-training on publicly available repositories with vulnerable code. Moreover, developers are called the "weakest link in the chain" since they have very minimal knowledge of code security. Although existing solutions provide a reasonable solution to vulnerable code, they must adequately describe and educate the developers on code security to ensure that the security issues are not repeated. 
Therefore we introduce a multipurpose code vulnerability analysis system \texttt{SecRepair}, powered by a large language model, CodeGen2 assisting the developer in identifying and generating fixed code along with a complete description of the vulnerability with a code comment. Our innovative methodology uses a reinforcement learning paradigm to generate code comments augmented by a semantic reward mechanism. Inspired by how humans fix code issues, we propose an instruction-based dataset suitable for vulnerability analysis with LLMs. We further identify zero-day and N-day vulnerabilities in 6 Open Source  IoT Operating Systems on GitHub. Our findings underscore that incorporating reinforcement learning coupled with semantic reward augments our model's performance, thereby fortifying its capacity to address code vulnerabilities with improved efficacy.
\end{abstract}

\section{Introduction}
\label{1_introduction}

In today's digital age, cybersecurity and code vulnerabilities are critical concerns that require constant attention and proactive measures to mitigate potential risks \cite{executiveorder, jointadvisory}. Vulnerabilities as small as the Null Pointer Exception, often referred to as the \textit{Billion Dollar Mistake} \cite{eichholz2019avoid} can adversely create security weaknesses and gaps that cyber criminals could exploit to propagate malware and ultimately carry out nefarious activities \cite{cisaalerts}. Code vulnerabilities can arise from technical glitches, human errors, Open-Source Software (OSS) reuse, and even unforeseen zero-day attacks \cite{woo2022movery, jang2012redebug, kim2017vuddy}.

\begin{figure}[H]
    \centering
    \includegraphics[scale=0.45]{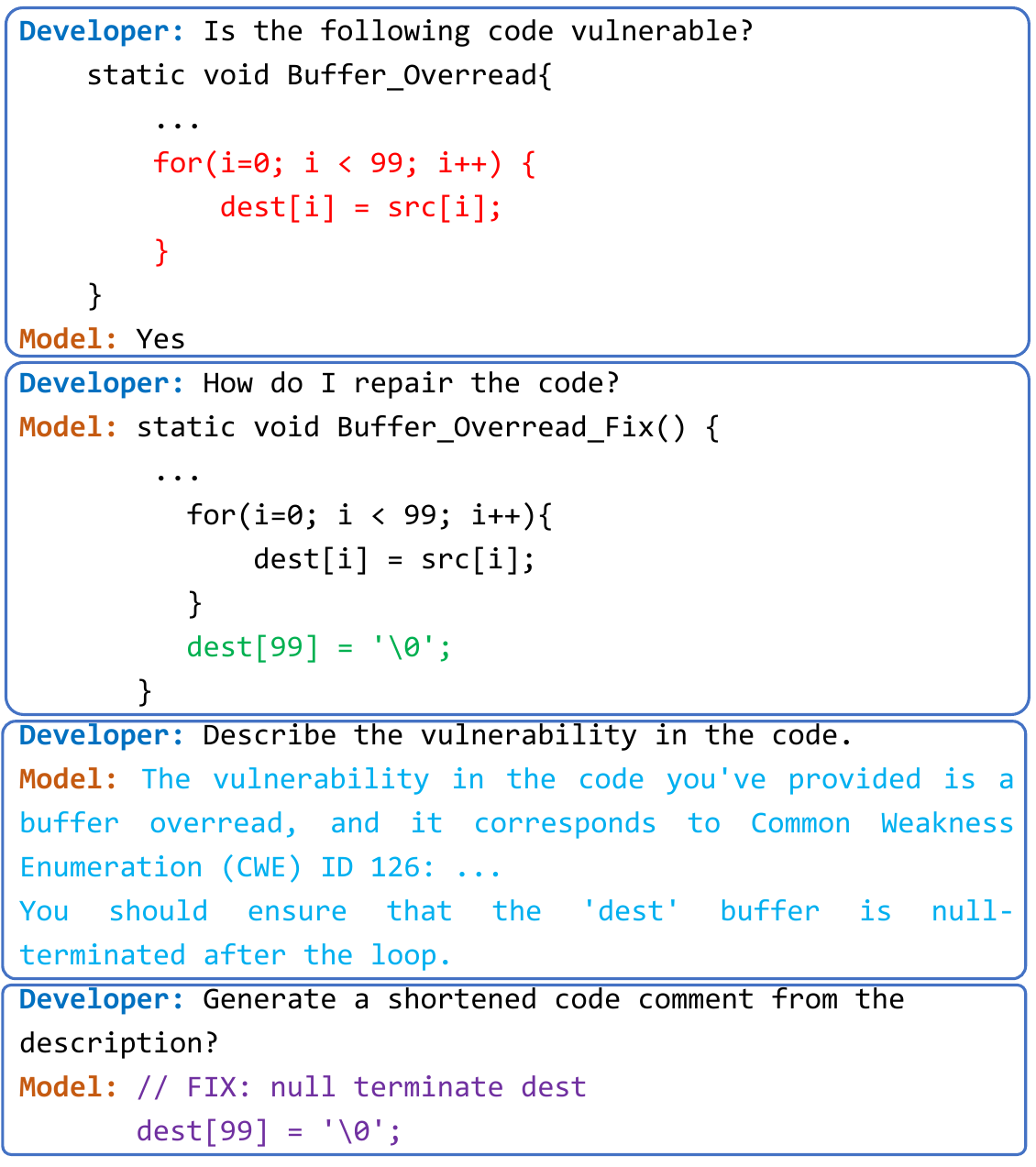}
    \caption{An illustrative example showcasing the input-output dynamics of our instruction-style model. We present a comprehensive depiction of our model's prowess across four distinct tasks: vulnerability identification, repair, description, and code comment generation.}
    \label{fig:fig_1}
\end{figure}

Recent advancements in neural language modeling and state-of-the-art Large Language Model (LLM)-based auto-completion tools, such as \texttt{GitHub Copilot} \cite{copilot}, \texttt{TabNine} \cite{tabnine}, \texttt{IntelliCode} \cite{intellij}, and \texttt{ChatGPT}, have significantly improved software architecture and development \cite{ahmad2023towards, murali2023codecompose, ross2023programmer}, making it a crucial feature in modern code editors and Integrated Development Environments (IDEs). However, these advancements have also raised valid questions about their training datasets and processes, generated code outputs, and most importantly, code semantic and security resilience.

Offering automatic, adequate, and comprehensive code vulnerability repair is the ultimate solution for the unprecedented and uncontrollable code vulnerabilities currently witnessed in software development. Security-oriented research has focused on detecting Vulnerable Code Clones (VCC). For example; \texttt{VUDDY} \cite{kim2017vuddy} proposes a fingerprinting mechanism to process and detect VCC in \texttt{Apache HTTPD} and \texttt{Ubuntu OS} distribution, \texttt{MVP} \cite{xiao2020mvp} addresses the problem of VCC by detecting mismatches between vulnerability and patch function signatures, \texttt{Movery} \cite{woo2022movery} presents a similar approach while having a superior capability in detecting internal VCC modifications. Furthermore, several other studies, including \texttt{VulRepair} \cite{fu2022vulrepair} and \cite{pearce2022examining}, have investigated the use of LLMs for automated code vulnerability repair. 



Unfortunately, despite their scientific innovation and potential benefits, each of the aforementioned systems has significant drawbacks and challenges in various aspects. These issues have hindered the widespread adoption of these solutions in developers' day-to-day workloads. Specifically, the challenges include: \textit{(i)} the failure to address security and vulnerability concerns during the code repair process; \textit{(ii)} reliance on limited or unreliable code signatures to detect mismatches; and \textit{(iii)} lack of comprehensive and transparent security and vulnerability measures during the repair process, including the failure to provide systematic commit messages and code vulnerability descriptions.

To address the critical need for identifying code vulnerability code and provide repairs to developers on vulnerable code, we are the first to introduce a new reliable AI-assisted solution \texttt{SecRepair} for vulnerability identification,  repair, and description with code comment generation. Our system leverages the power of CodeGen2 \cite{nijkamp2023codegen2}, a large language model designed and explicitly fine-tuned for code security analysis to identify and repair vulnerable code. To facilitate developers with a comprehensive security analysis, we curate an extensive instruction-based dataset \texttt{InstructVul} encompassing diverse prompts tailored to security concerns of source code vulnerabilities. In order to generate commit comments for the developers, we provide a reinforcement learning technique with semantic reward to provide a concise and appropriate code commit comment. Figure \ref{fig:fig_1} shows an overall depiction of our contribution to vulnerability analysis.


In summary, the contributions of this paper are:
\begin{itemize}
    \item We introduce a state-of-the-art AI-assisted tool, \texttt{SecRepair} designed to identify vulnerable codes and repair them with a customized description generation suited for developers with limited security knowledge. Our system also leverages Reinforcement Learning with semantic reward value to enhance its capabilities and effectiveness in generating comments for code commits.

    \item To support our systems training process and prepare a robust model for vulnerability analysis, we are the first to have compiled and released a comprehensive instruction-based vulnerability dataset \texttt{InstructVul}. Our dataset is specifically tailored to the C/C++ programming language, emphasizing its significance in terms of security vulnerabilities. 

    \item Our research also includes a quantitative analysis of the results with zero-day and N-day analysis along with ablation studies and case studies qualitatively and quantitatively demonstrate the quality of our model. 

\end{itemize}


\section{Related Work}
\label{2_related_work}

\paragraph{\textbf{Code Vulnerability with LLMs.}} With the rapid progress of Large Language Models (LLMs), their application in code development-related tasks has significantly increased. GitHub Copilot \cite{copilot} has initiated a trend towards AI-assisted software development \cite{ahmad2023towards, murali2023codecompose, ross2023programmer}. Notably, Codex \cite{chen2021evaluating} introduced a docstring-based approach where developers input instructions as docstrings, and LLMs generate corresponding code. However, most Large Language Models (LLMs) are less concerned about the security issue that comes with programming languages. Pearce \cite{pearce2022examining} examined LLMs' zero-shot vulnerability repairing capabilities and the bugs' challenges. Their experimental results showed that, although LLMs can generate bug fixes, they need a specifically constructed prompt to fix a particular bug. SVEN \cite{he2023controlling} proposed an adversarial technique to evaluate LLMs. They propose to generate safer code by leveraging property-specific continuous vectors to guide program generation. Moreover, \cite{pearce2022asleep},  \cite{jesse2023large} and \cite{sandoval2022security} provided an extensive study on the effectiveness of autocompletion by integrating them with different IDEs and analyzing their outcome. Although \cite{pearce2022asleep},  \cite{jesse2023large} concluded that LLMs do produce vulnerable code, Sandoval \cite{sandoval2022security} suggested that LLMs help in generating more functional codes with improved security bugs.

\paragraph{\textbf{Vulnerability Repair.}} Generating repair of programs is a challenging task because to repair a program, initially, we need the vulnerable line and, furthermore, generate an appropriate compilable line to fix the vulnerability. Some earlier works to generate vulnerability patches involve human-specified safety properties \cite{huang2019using}. For example, not allowing a program to access memory beyond its boundary. Similar works proposed by Zhang et al. \cite{zhang2022program} proposed to find patch invariants or vulnerable locations of code and use a set of templates in order to generate a repair. More recent works like Vrepair \cite{chen2022neural} include a transformer-based, transfer learning approach to fix vulnerabilities in real-world programs. Similarly, VulRepair \cite{fu2022vulrepair} proposed a vulnerability repair technique using pre-trained CodeT5 \cite{wang2021codet5} and BPE tokenizer. However, with the advent of code-based LLMs like Codex \cite{chen2021evaluating}, works proposed by Pearce et al. \cite{pearce2022examining} and Jesse et al. \cite{jesse2023large} and Prenner \cite{prenner2021automatic} demonstrated their capability to repair vulnerable code with zero-shot learning.

\begin{figure*}[h!]
    \centering
    \resizebox{\linewidth}{!}{
    \includegraphics[]{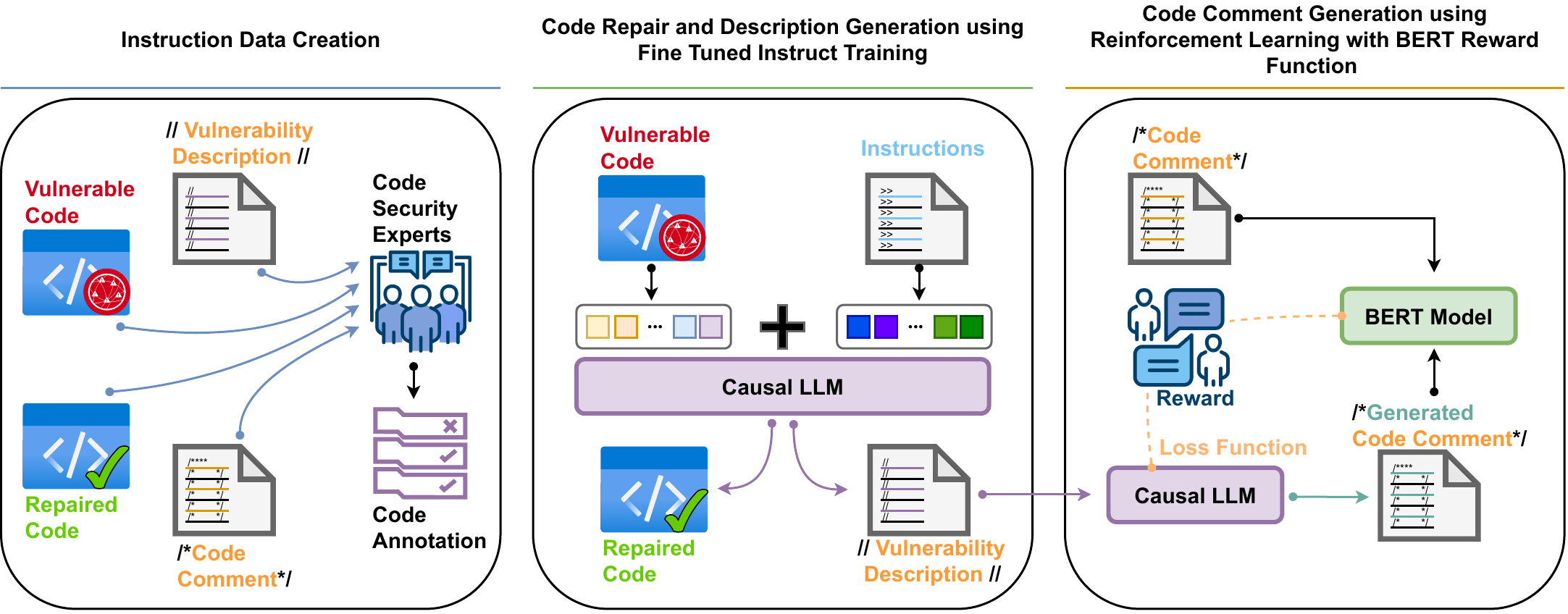}
    }
    \caption{The overall architecture of SecRepair includes (1) Instruction Dataset Preparation: Creating a triplet of instruction, context input and output, (2) Code Repair and Detection: Trains the model for vulnerable code identification, repair, and description, (3) Fine Tuning using Reinforcement Learning: Reinforcement learning with semantic reward to further fine-tune the model for code comment generation.}
    \label{fig:fig_2}
\end{figure*}


\section{Approach}
\label{3_methodology}

This work introduces a code vulnerability repair system \texttt{SecRepair}, powered by LLM, and incorporates reinforcement learning with a unique semantic reward technique. The proposed system is designed to help developers generate fixed code while comprehensively describing the vulnerability with a code comment. Figure \ref{fig:fig_2} illustrates the architecture of our proposed work. Specifically, we divide the task of code vulnerability repair into two main stages: \textit{(i)} code repair and vulnerability description generation; and \textit{(ii)} code comment generation.

\subsection{Code Vulnerability Repair and Description}

To achieve code vulnerability repair and description, a prior fundamental step needs to be completed, namely, vulnerability identification and localization (i.e., pinpointing the vulnerable line in the code). This fundamental step, along with the proposed pipeline, heavily relies on a carefully-created instruction dataset consisting of vulnerable code coupled with its repaired code (discussed in \textit{Instruction Dataset}). 

\textbf{Code Vulnerability Repair.} At stage two of Figure \ref{fig:fig_2}, there are two data inputs: a vulnerable code and an instruction to train SecRepair. The model learns the representation of the code through the training process, enabling vulnerability localization and repair. Once the model has been adequately trained on instruction fine-tuning for code vulnerability detection, it inherently achieves representation for localization. Our dataset provides repaired versions for each vulnerable code, facilitating our model's understanding of the difference between vulnerable and non-vulnerable code pairs. This method imparts an internal representation of localized vulnerable lines during vulnerability identification training. Additionally, the model is optimized for vulnerable code detection and repair by using Cross-Entropy Loss as the optimization criterion, and beam search \cite{freitag2017beam} is utilized to select multiple vulnerability repair candidates based on conditional probability for our input.

One of the major challenges with code vulnerability repair is that the code sample can be composed of a high number of tokens. Moreover, code writing, execution, and repair is a sequential task that requires a sequential understanding of the code tokens. However, regular encoder-decoder-based LLM architecture shows a bidirectional property where they read tokens from forward and backward. Such a bidirectional architecture causes unnecessary computation that is not needed for code vulnerability repair. Therefore, we introduce a simple but effective modification to encoder-decoder architecture for longer code sequences by dropping the encoder module. Then we combine the input and output sequences into a single sequence and train as a standard language model. In addition to faster inference time, we also decrease the memory footprint by almost half. Thus, we convert the entire input-output sequence into a sequence of tokens, $t_1, t_2, ... t_p \in T$ and $y_1, y_2, ... y_q \in Y$, into a single  sequence $w_1, w_2, ... w_{p+q}$ = $(t_1, ..., t_p, \$,  y_1, ... , y_q)$, where $\$$ is a special token separating inputs and outputs. Here $p$ is the number of input tokens, and $q$ is the number of output tokens. We train the model auto-regressively to predict the next word given the previous ones using the following equation:
\begin{equation}
    p(w_1, w_2, ... w_{p+q}) = \Pi_{j=1}^{p+q}   p(w_i|w_1, ..., w_{j-1})
\end{equation}

Given the causal decoder architecture of \texttt{SecRepair}, our model is forced to predict the next code token, and the model cannot overlook future tokens by looking at the next token during output generation. The model is provided with the input sequence $t_i \in T$  during inference, which in turn auto-regressively generates the output, $y_i$.

\textbf{Code Vulnerability Description.} At stage 2 of Figure \ref{fig:fig_2}, we also fine-tune our model as a code-to-text-based format to generate a developer-friendly code descriptions. However, vulnerability description is different from code repair because it requires a sequential understanding of the code and a sequential generation of the vulnerability description. The causal decoder architecture is optimized with the auto-regressive property of our model  to consider the context of the already generated tokens, therefore ensuring the sequential generation of code descriptions.


\subsection{Reinforcement Learning for Code Comment}


At stage 3, denoted in \ref{fig:fig_2} the pre-trained SecRepair is cloned and employed to generate a developer-friendly description. After the successful generation of developer-friendly descriptions, we generate code comments. Code comments presents shorter version of a developer-friendly description while maintaining the same semantic meaning. It is a concise version of the description which the developers place at the code repair location as a commit message. Therefore, we fine-tune our model for a text-to-text task where the generated text is significantly shorter but still conveys the semantic meaning of the original description. To achieve the latter, we propose a reinforcement learning technique with a semantically aware reward function with the Proximal Policy Optimization (PPO) algorithm outlined by Schulman et al. \cite{schulman2017proximal} to fine-tune the learning environment. We denote the input as the developer-friendly description, $D$, and the commit comment as $D_c$, and the output sequence is $w^c_1, w^c_2, ... w^c_k$, where $k$ is the total number of output tokens which is significantly shorter than the length of $D$. We define the token-wise generative summarization process as a deterministic Contextual Markov Decision Process \cite{hallak2015contextual}  with observable context from previous tokens. The sequence generated is the state at the $k^{th}$ token generation. A policy $\pi (. | w^c:k-1, t)$ is the probability distribution over all input tokens from description $D$, conditioned on the context and state. 

\textbf{Policy Optimization.} The reinforcement learning objective is to find the optimal policy that maximizes the cumulative reward signal by shortening the description and by keeping the semantic meaning. To meet this requirement, we introduce a BERTScore \cite{zhang2019bertscore} with reinforcement learning as a metric that quantifies the semantic similarity between two texts and generates a reward score. If the original description is $D$, commit message description generated by the model is $\hat{D_c}$ and the ground truth commit comment is $D_c$. As such, we calculate the policy optimization $r_\theta$, using the following equation:
\begin{equation}
    \label{eqn:3}
    L(r_\theta) =  log (\sigma(r_\theta (D, D_c) - r_\theta (D, \hat{D_c}) )
\end{equation}

where $r_\theta (D, D_c)$ and $r_\theta (D, \hat{D_c})$ is the scalar output of the reward model for the description $D$. Here $\sigma$ is the activation function, and $\theta$ is the learnable parameter.



\textbf{Reward.} Reward is calculated by introducing BERTScore, a semantic comparison using the cosine similarity score. A BERT vector representation of tokens permits the generation of a soft measure of similarity rather than exact matching. The cosine similarity of a reference token $d_C^i$ and a candidate token $\hat{d_C^i}$, we calculate the cosine similarity as $(d_C^i)^T \hat{d_C^i}$. Therefore the F1 measurement of the BERTScore stands as follows; 

\begin{equation}
    R_{BERT} = \frac{1}{|d_C^i|} \sum_{d_C^i \in d} max (d_C^i)^T \hat{d_C^i}
\end{equation}

where $R_{BERT}$ is our expected BERTScore.


\section{Instruction Dataset}
\label{4_instruct_dataset}

This work is the first to introduce \texttt{InstructVul}, a new instruction-based dataset for vulnerability identification and repair system with a description and code comment generation. \texttt{InstructVul} is comprised of four components based on four tasks; they are 1) Vulnerability Identification, 2) Vulnerability Repair, 3) Vulnerability Description, and 4) Code Comment Generation.  The following subsections define our dataset formally, followed by an in-depth description of its creation.

\subsection{Formal Definition}

Each data entity contains three components: instruction, context input, and output. The instruction is denoted by $I$, which is a statement stating one of the four tasks. The second component context input $C_i$ is a supplement for each instruction. The third component is the output $y$, the expected answer we want our LLM to generate.

\subsection{Dataset Creation}
To create an instruction-based dataset for identification we initially leverage the vulnerable and non-vulnerable function pairs from VulDeelocator \cite{li2021vuldeelocator}, which consists of multiple C/C++ files, each containing a vulnerable code and corresponding repaired code. Building upon this foundation, we extend the dataset by incorporating instructions with descriptions and code comments.

\paragraph{\textbf{Input generation, ($I$, $C_i$).}}
Our input consists of the instruction we provide in natural language, and the context input is a code. We combine these two to generate the final input. Based on the four tasks we perform, our security experts created 20 seed questions for each task. In order to provide context to the instruction, we use the vulnerable code as the contextual input. Context input act as a supplement to the instruction. 


\paragraph{\textbf{Output Generation, ($y$).}}
Based on the task provided by our instruction, we provide four types of output. Firstly if the instruction asks to identify the vulnerable code, the output is \texttt{yes/no}, indicating the existence of a vulnerability in the code. Secondly, if the instruction asks about generating the fixed code, the output is the repaired code we obtained from VulDeelocator. If the instruction is about describing the vulnerability in code, the output will be a description of why the code is vulnerable with a possible hint to fix it. Finally, the if the instruction asks to generate a code comment, the output will be a shortened and concise version of the original description.

\hspace{5mm}\textit{Description Generation:} We consider code description as a combination of describing the code objective and explaining the vulnerability of the code with a possible hint to fix the vulnerability. We generate the code description using the code generative model CodeAlpaca \footnote{https://github.com/sahil280114/codealpaca}. This model was initially fine-tuned to generate texts based on instruction. Therefore we use this model to generate the objective of vulnerable code as a part of our description. We also extract the generic description of the vulnerable code for CWE \cite{cwe} with a hint to solve the vulnerability. We combine these two together and create a comprehensive description for each vulnerable code.

\hspace{5mm}\textit{Code Comment Generation:} Furthermore, we added code comments for each vulnerable code. A code comment is a shortened and concise version of the vulnerable code that a developer uses during a code commit. Our security experts and developers handcrafted 1000 code commits which took approximately 80 human hours. They used the descriptions as a suggestion to generate the code comments. In order to keep the comments concise, all the comments are kept between one to three lines.

After creating all the components, we combine these with specific instructions based on their task. Therefore in total, we have 18086 entities in our dataset.

\subsection{Quality of Data}
Extracting vulnerable and repaired functions from a single file in the VulDeelocator dataset presented challenges due to inconsistent formatting. The placement of vulnerable code and their repaired code were always not consistent. Furthermore, approximately in 5\% of cases, vulnerable code does not have a fix and was not added as a part of our dataset. We developed a robust regular expression to identify each function's start accurately and used JOERN \footnote{https://joern.io/} to extract individual vulnerable and non-vulnerable code. Subsequently, we verified the correctness of the repaired code manually by having security experts randomly examine 200 entities of our dataset with their corresponding repairs, descriptions, and code comment.

\section{Experiments and Discussions}
\label{5_experiments}

In this section, we will go over the evaluation metrics and then provide details on our experimentation based on the research questions we provided.

\subsection{Evaluation Metrics}
Since we use a generative model for vulnerability repair, description generation, and code comment, we used the metrics like BLEU, Rouge-L, and human evaluation scores suitable for generative models. However, we use F1, Precision, Recall, and Accuracy for vulnerability identification tasks.

\paragraph{\textbf{BLEU Score:}}The BLEU \cite{papineni2002bleu} score is a syntax-based way for evaluating machine-generated text between 0 and 1. It is a reference-based metric and may not capture all aspects of translation quality, such as fluency or semantic accuracy.

\paragraph{\textbf{Rouge-L:}}Similar to BLEU, Rouge-L \cite{lin2004rouge} score is also a number between 0 and 1 to measure the syntax-based similarity of two generated texts. It generates a score by quantifying precision and recall by examining the longest common subsequence (LCS) between the generated and reference codes.

\paragraph{\textbf{Human Evaluation:}}In order to determine how the generated outcome helps the developers and security experts, we provide a human evaluation by our internal developers and security experts to analyze how the outcome has helped them solve real-world problems.

\subsection{Results and Discussions}
For all of our experiments, we randomly divided our datasets into 80:10:10 for training, validation, and testing. We used a pretrained CodeGen2 model with 32 layers of decoders. We trained the model for three epochs with a maximum token length of 512, and our learning rate was $2e^{-5}$ with a batch size of 2 for our 7B parameter model. We also use a beam size of 4 for the generation task and a temperature value of 0.5 for optimal performance. We used 8 NVIDIA A100 GPUs, each with 40 GB of memory. Furthermore, we integrated DeepSpeed \cite{rajbhandari2020zero} with our HuggingFace implementation of CodeGen2 to reduce memory footprints. During training, our model uses approximately 15 GB of memory per GPU for an optimized training environment with the minimum inference time.

\begin{table}[b]
\centering
\begin{tabular}{p{0.15\textwidth}
                 p{0.03\textwidth}
                 p{0.025\textwidth}
                 p{0.025\textwidth}
                 p{0.025\textwidth}
                 p{0.025\textwidth}}


\toprule
\textbf{Model} & \multicolumn{1}{l}{\textbf{Parameters}} & \textbf{F1}   & \textbf{Pre.} & \textbf{Rec.} & \textbf{Acc.} \\ \midrule
Devign         & $<$1M                                       & 0.56          & 0.55          & 0.55          & 0.56          \\
VELVET          & $<$1M                                       & 0.62          & 0.61          & 0.59          & 0.68          \\
PFGCN          & 110M                                       & 0.64          & 0.64          & 0.61          & 0.62          \\
CodeT5        & 770M                                    & 0.68          & 0.62          & 0.59          & 0.68          \\
CodeGen2        & 1B                                      & 0.72          & 0.70          & 0.68          & 0.79          \\
CodeGen2       & 3.7B                                    & 0.75          & 0.77          & \textbf{0.73} & 0.85          \\
\textbf{SecRepair}      & 7B                                      & \textbf{0.82} & \textbf{0.80} & 0.70          & \textbf{0.88} \\ \bottomrule
\end{tabular}
\caption{Comparison of our Model with SOTA approaches in identifying vulnerabilities on our instruction dataset InstructVul.}
\label{tab:tab1}

\end{table}


In this study, we aim to address three research questions (RQs) concerning the effectiveness and capabilities of our proposed system for vulnerability analysis.

\begin{itemize}
    \item \textbf{RQ1:} Can we automatically identify code vulnerability and accurately repair the code? 
 
    \item \textbf{RQ2:} Can we comprehensively describe the code vulnerability of the proposed repair to the developer?
  
    \item \textbf{RQ3:} Can we optimize and summarize the program description and generate code comments?
\end{itemize}

In the remaining part of this section, we will conduct the experiments by answering the three research questions. \\

\indent\textbf{RQ1: Can we automatically identify code vulnerability and accurately repair the code? }

To ensure accurate vulnerability repair, a robust system should identify vulnerabilities accurately. To evaluate the effectiveness of our vulnerability identifier, we conducted experiments by training our model on our proposed dataset. Furthermore, the goal of the vulnerability repair system is to provide a reasonable repair of the original vulnerable code. However, depending on the developer and the application's business logic, a vulnerable code may be repaired in various ways. Therefore, in the case of vulnerability repair tasks, we use BLEU and Rouge-L metrics.

\textbf{Discussion.} For vulnerability identification, the aim was to demonstrate our model's superior vulnerability identification capability in vulnerability detection tasks compared to existing models. In the first stage of our experiments, as shown in Table \ref{tab:tab1}, our model outperformed other state-of-the-art models. Our model achieved a notable improvement in F1 score of 7\% compared to CodeGen-3.7B and almost a 14\%  improvement compared to CodeT5.

\begin{table}[b]
\centering

\begin{tabular}{@{}llll@{}}
\toprule
\textbf{Model}                 & \textbf{Parameter} & \textbf{BLEU} & \textbf{Rouge-L} \\ \midrule
StarCoder                      & 1.1B               & 0.17          & 0.13             \\ \midrule
\multirow{2}{*}{CodeT5}        & 2B                 & 0.64          & 0.71             \\
                               & 6B                 & 0.73          & 0.79             \\ \midrule
\multirow{2}{*}{CodeGen-Multi} & 2B                 & 0.54          & 0.57             \\
                               & 6.7                & 0.80          & 0.98             \\ \midrule
\multirow{3}{*}{\textbf{SecRepair}}      & 1B                 & 0.42          & 0.31             \\
                               & 3.7B               & 0.74          & 0.91             \\
       & 7B                 & \textbf{0.82} & \textbf{0.98}    \\ \bottomrule
\end{tabular}
\caption{BLEU and Rouge-L score on our dataset with models trained on code. We use LLMs with multiple parameters to compare our work.}
\label{tab:tab2}

\end{table}

Our experimental results demonstrate the superiority of our model compared to other models with similar or fewer parameters. In Table \ref{tab:tab2}, we observe that model achieves a 0.09  BLEU higher than the 6B variant of CodeT5 and 0.02 BLUE higher for the 6.7B variant of CodeGen-Multi. Moreover, the 1B variant of StarCoder, SecRepair, exhibits notably lower performance when compared to models with parameter counts equal to or exceeding 3B. An intriguing observation from the comparison of BLEU and Rouge-L scores is that, in most instances, the Rouge-L score remains consistently higher. This phenomenon arises because the BLEU score evaluates similarity based on n-grams and compares the match between those n-grams. In contrast, the Rouge-L score accurately measures similarity using the least common sequence (LCS) between the original and generated output tokens, providing a more reliable evaluation of the generated code.

\textbf{Wild Testing.} Furthermore, we also provide wild testing by demonstrating the vulnerability identification capability of our system on six IoT device OS, namely Contiki, Zephyr, FreeRTOS, RIOT-OS, and Raspberry Pi. Initially, we scanned the repositories using JOERN to extract the functions from IoT operating systems and detected 30 n-day vulnerabilities, and our security experts further validated the results. The confirmed zero-day vulnerabilities were associated with two vulnerable instances in Zephyr, two in RIOT-OS and one in FreeRTOS.

\begin{table}[t]
\centering
\begin{tabular}{@{}lllll@{}}
\toprule
\textbf{Model}                                                              & \textbf{Parameter} & \textbf{BLEU} & \textbf{Rouge-L} & \textbf{Human} \\ \midrule
CodeT5                                                                      & 770M               & 0.34          & 0.39             & 2              \\ \midrule
\multirow{3}{*}{CodeGen}                                                   & 1B                 & 0.45          & 0.87             & 3              \\
& 3.7B               & 0.71          & 0.70             & 3              \\
& 7B                 & 0.68          & 0.94             & 4              \\ \midrule
\multirow{3}{*}{\begin{tabular}[c]{@{}l@{}}SecRepair \\ \end{tabular}} & 1B                 & 0.57          & 0.90             & 3              \\
& 3.7B               & 0.47          & 0.72             & 4              \\
& 7B                 & 0.76          & 0.98             & 5              \\ \bottomrule
\end{tabular}
\caption{BLEU, Rouge-L, and Human Evaluation on developer-friendly vulnerability description task.}
\label{tab:tab3}

\end{table}


\indent\textbf{RQ2: Can we comprehensively describe the code vulnerability of the proposed repair to the developer?}

While vulnerability repair provides a plausible solution to vulnerable code, a proper solution depends on whether the repair agrees with the business logic or whether the developer understands the vulnerability and repair. Therefore to complement vulnerability repair, we further provide the developers with a cybersecurity description of the vulnerability.


\textbf{Discussion} Table \ref{tab:tab3} compares different variants of our model with other code-to-text generative models like CodeT5 and CodeGen models with different parameters.
Furthermore, we also evaluate the outcome based on evaluations from our internal security experts and developers. To bring them into the evaluation process, we provide them with 20 sample vulnerable code, their original descriptions, and the description generated by the model. Therefore for each vulnerable code, each developer gets 7 outcomes to compare with the original outcome and average the outcomes. We ask the developers to provide a number between 1 to five, where 1 means they are least satisfied with the description generated by the model, and 5 represents the developers are completely satisfied with the description.

Table \ref{tab:tab3} shows an improvement of 0.08 BLEU score and 0.04 Rouge-L score when we compare the 7B variant of CodeGen and SecRepair model. Similarly, we see an improvement for CodeT5 when the model. Furthermore, we see that the average human scores also increase from 4 to 5 when we compare the outputs from the 7B variant of our SecRepair and CodeGen model.

\indent\textbf{RQ3: Can we optimize and summarize the program description and generate code comments?}

In this experimental study, create a clone of our last trained model from RQ2, and use it as a text-to-text model using reinforcement learning to generate a concise text while preserving the semantic meaning of the original description. Our goal in this experiment is to demonstrate that our model, when trained using reinforcement learning with semantic reward, performs superior to regular instruction-based fine-tuning using a cross-entropy loss.

\textbf{Discussion.} Table \ref{tab:tab4} compares different variants of our model with instruction-based training and reinforcement learning-based training. Here FT denotes regular instruction fine-tuning, and RL denotes reinforcement learning. Similar to RQ2, we measure the efficacy of our model using BLEU and Rouge-L and human evaluation. Table \ref{tab:tab4} shows an improvement of 0.06 BLEU score and 0.05 Rouge-L score when we compare the 7B variant of CodeGen and our SecRepair model. Furthermore, the human score also improves when using a semantic reward. Similarly, we see an improvement in the 1B and 3.7B variants when trained using reinforcement learning. The consistent improvement in reinforcement learning clearly shows the efficacy of our training using a semantic reward value.

\begin{table}[t]
\label{tab:tab4}
\centering
\begin{tabular}{@{}lllll@{}}
\toprule
\textbf{Model}                                                              & \textbf{Parameter} & \textbf{BLEU} & \textbf{Rouge-L} & \textbf{Human} \\ \midrule

\multirow{3}{*}{\begin{tabular}[c]{@{}l@{}}SecRepair \\ (FT) \end{tabular}} & 1B                 & 0.57          & 0.90             & 3              \\
& 3.7B               & 0.47          & 0.72             & 3              \\
& 7B                 & 0.54          & 0.67             & 3              \\ 
\midrule

\multirow{3}{*}{\begin{tabular}[c]{@{}l@{}}SecRepair \\ (RL) \end{tabular}} & 1B                 & 0.50          & 0.65             & 3              \\
& 3.7B               & 0.53          & 0.70             & 4              \\
& 7B                 & 0.60          & 0.72             & 5              \\ \bottomrule
\end{tabular}
\caption{BLEU, Rouge-L, and Human Evaluation on generating code comments.}
\label{tab:tab4}
\end{table}

\section{Ablation Studies}
\label{6_ablation}
    

\begin{figure}[hb]
    \centering
    \includegraphics[scale=0.46]{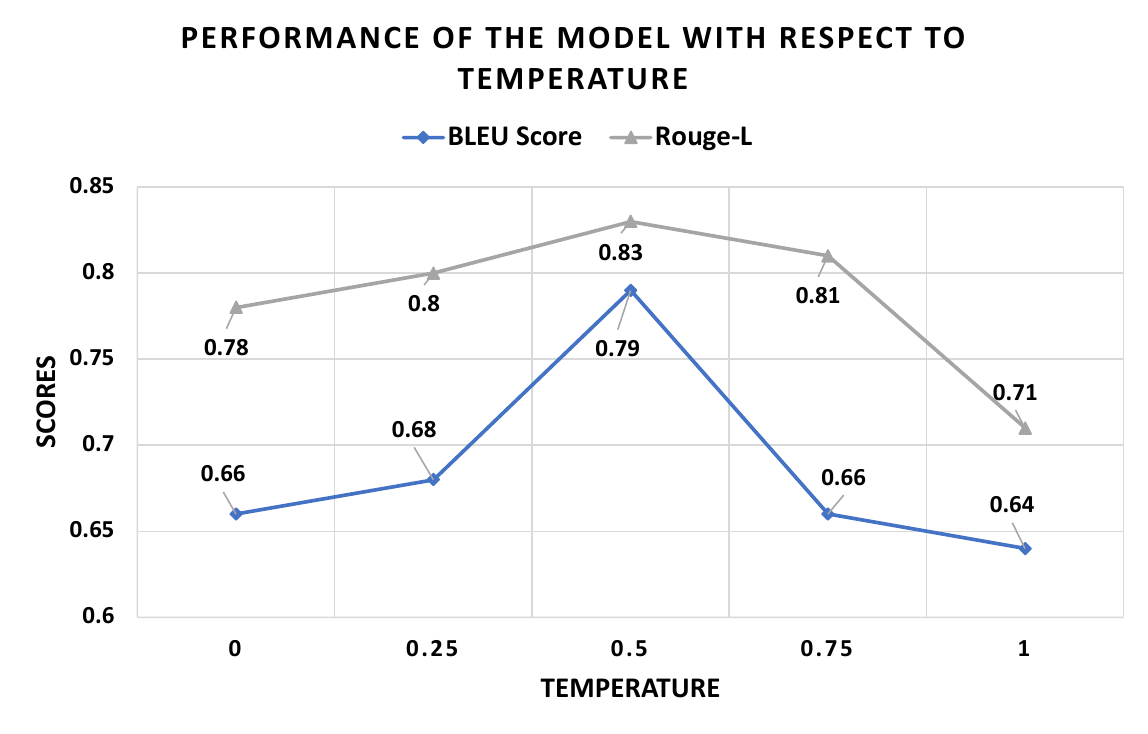}
    \caption{Ablation Study on the Performance of our Model with varying temperature}
    \label{fig:fig_3}
\end{figure}



This section explores the impact of two crucial components, temperature and beam size, on generative models. Higher temperature leads to token generation with reduced probabilistic values. Meanwhile, increasing the beam size in beam search enables tracking the top-k probable sequences and extending them with next-token probabilities.


\textbf{Temperature:} Figure \ref{fig:fig_3} shows how the beam BLEU and Rouge-L score change with the change in temperature.  In order to test the effect of temperature, we kept the beam search at its default value and tested the outcome with temperature values of 0.0, 0.25, 0.50, 0.75, and 1.0. We observe in Figure \ref{fig:fig_3} that the scores are the lowest when the temperature is 0.0 and begin to increase when the value is at its sweet spot at 0.50. However, we again see a decrease in the score when the temperature is close to 1.

\begin{figure}[t]
    \centering
    \includegraphics[scale=0.49]{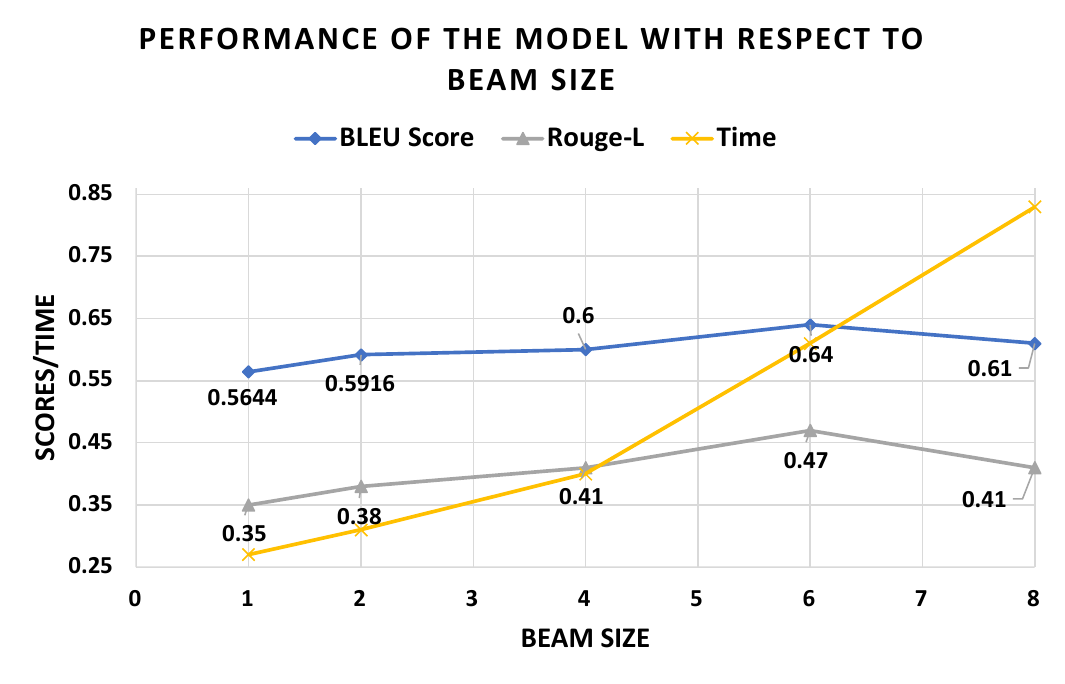}
    \caption{Ablation Study on the Performance of our Model with Varying Beam Size}
    \label{fig:fig_4}
\end{figure}

\textbf{Beam Search:} Similarly, in Figure \ref{fig:fig_4}, when we keep the temperature value constant and increase the beam value to 1, 2, 4, 6, and 8. We see slight improvements in the results as the beam size increases. However, this improvement comes with some drawbacks. The yellow line shows the relative time converted to a range between 0 and 1 for convenience. We see that the inference time increases almost exponentially with the increase in beam values. Here, we converted the time values within the range of 0 to 1 for the convenience of displaying them in a single graph. While the increase of beam value improves the performance, higher inference time makes this an inconvenience for the developers to use in real-time.

\section{Conclusion}
\label{8_conclusion}

Our study presents a comprehensive end-to-end solution \texttt{SecRepair}, to address the challenges associated with code vulnerabilities with code generated from LLM and OSS code. We introduced an instruction-based dataset to train our model SecRepair powered by CodeGen2-7B, to address the challenge of source code vulnerability identification, repair, and description. We used reinforcement learning with semantic reward to further fine-tune our model to provide code comments. Since the generated outputs do not follow a specific formatted output, we use BLEU and Rouge-L scores to measure the outcome of our model. We further tested the generative capability of our model using human evaluation for developer-friendly description and code comment generation tasks. Our proposed model demonstrates superior scores in identifying, repairing, and describing vulnerabilities with code comments with an ablation study on the effect of temperature and beam search.


\bibliography{aaai24}

\end{document}